%% file: iclr2025_conference.tex
\documentclass{article} 
\usepackage{iclr2025_conference,times}
\usepackage{booktabs}
\usepackage{subcaption}
\input{math_commands.tex}

\usepackage{pifont}
\usepackage{hyperref}
\usepackage{url}
\usepackage{graphicx}
\usepackage{tabularx} 
\usepackage{multirow}
\title{Life-Cycle Routing Vulnerabilities of LLM Router} 

\iclrfinalcopy
\author{Qiqi Lin$^{1}$\thanks{Equal contribution}\, , Xiaoyang Ji$^{2}$\footnotemark[1]\, , Shengfang Zhai$^{1}$\footnotemark[1]\;\,\thanks{Corresponding authors.}\, , Qingni Shen$^{1}$, Zhi Zhang$^{3}$, Yuejian Fang$^{1}$\footnotemark[2]\\
~\textbf{Yansong Gao}$^{3}$\\
$^1$School of Software and Microelectronics, Peking University\\
$^2$College of Cyber Science, Nankai University \\
$^3$The University of Western Australia \\
\texttt{\{qiqilin,zhaisf\}@stu.pku.edu.cn},     
\texttt{\{qingnishen,fangyj\}@ss.pku.edu.cn} \\
\texttt{\{jixiaoyang666,zzhangphd\}@gmail.com}, \;
\texttt{garrison.gao@uwa.edu.au}
}

\begin{document}

\maketitle

\begin{abstract}
Large language models (LLMs) have achieved remarkable success in natural language processing, yet their performance and computational costs vary significantly. LLM routers play a crucial role in dynamically balancing these trade-offs. While previous studies have primarily focused on routing efficiency, security vulnerabilities throughout the entire LLM router life cycle, from training to inference, remain largely unexplored. In this paper, we present a comprehensive investigation into the life-cycle routing vulnerabilities of LLM routers. We evaluate both white-box and black-box adversarial robustness, as well as backdoor robustness, across several representative routing models under extensive experimental settings. Our experiments uncover several key findings: 1) Mainstream DNN-based routers tend to exhibit the weakest adversarial and backdoor robustness, largely due to their strong feature extraction capabilities that amplify vulnerabilities during both training and inference; 2) Training-free routers demonstrate the strongest robustness across different attack types, benefiting from the absence of learnable parameters that can be manipulated. These findings highlight critical security risks spanning the entire life cycle of LLM routers and provide insights for developing more robust models.

\end{abstract}

\section{Introduction}

In recent years, large language models (LLMs) such as GPT-3.5 
\citep{brown2020language}, GPT-4 \citep{achiam2023gpt}, and PaLM 2 \citep{anil2023palm} have achieved significant progress in natural language processing tasks, finding widespread applications in open-domain dialogue, question answering, code generation, and other tasks \citep{gu2023llm,zhuang2023toolqa,ghosh2024clipsyntel}. 
However, different LLMs vary in terms of training data, model size, and computational cost, leading to differences in their strengths, weaknesses, and overall capabilities. Generally, larger models tend to exhibit stronger performance but come with higher inference costs, whereas smaller models are more computationally efficient but have limited capability in handling complex tasks.

LLM Routing \citep{ding2024hybrid,ong2024routellm,hu2024routerbench} is a state-of-the-art optimization strategy designed to mitigate this trade-off and achieve a balance between response quality and computational cost.  Differing from reward modeling \citep{ouyang2022training}, LLM routers can dynamically select the appropriate model among multiple LLMs based on the difficulty of queries. This approach reduces computational overhead while maintaining performance, eliminating the need to retrain new LLM models for specific tasks. Due to these advantages, we have already witnessed various real-world applications, such as the emergence of Unify \citet{unifyai} and \citet{martian}.

Nevertheless, existing research primarily focuses on improving the performance and efficiency of routers. Non-predictive routing methods \citep{jiang2023llm,chen2023frugalgpt,aggarwal2023automix} adopt ensemble or cascading approaches to select the appropriate model based on the response quality of multiple models. Although these methods alleviate the cost-performance trade-off to some extent, they rely on multiple LLM queries, which may lead to increased latency. To more effectively reduce latency, recent studies \citep{ding2024hybrid, chen2024RouterDC, feng2024graphrouter} focus on predictive routing methods. \citet{lu2023routing} is the first to propose dynamically selecting the most suitable expert LLM using a pretrained reward model. \citet{ding2024hybrid} introduces a hybrid inference approach, while \citet{chen2024RouterDC} further incorporates a query routing method based on dual contrastive learning to leverage model advantages. Based on these advancements, RouteLLM \citep{ong2024routellm} introduces additional routing strategies, including similarity weighting and matrix factorization. These methods extract the complexity features of user queries to select the appropriate model before language model inference.
Despite significant progress in performance optimization, the security of predictive routers remains unexplored.  
This raises several critical research questions regarding the security risks across the entire life cycle of LLM routers:
\begin{itemize}
\vspace{-0.25cm}
    \item Q1 (Inference-time Attacks): Can a deployed LLM router be manipulated by adversaries to misroute queries, deliberately selecting larger models to cause resource consumption? 
    \item Q2 (Training-time Attacks):
    Can an LLM router be implanted with a backdoor during the training phase, allowing specific model selections to be triggered by predefined inputs?
    \item Q3 (Lifecycle-based Vulnerability Analysis): What are the differences in robustness among different types of router architectures at various stages of the router's life cycle?
\vspace{-0.25cm}
    
\end{itemize}

To address these concerns, we systematically investigate the life-cycle vulnerabilities of LLM routers. In this paper, we identify two practical attack surfaces introduced by the router module across its life-cycle stages:

\ding{182} Adversarial Attacks (Q1): For deployed router models, there are two types of attack scenarios based on whether the router is open-source: white-box and black-box attacks. In these scenarios, attackers extract gradient or sample information to construct malicious universal triggers, using adversarial attack methods to force the router to select a specific LLM for query response.

\ding{183} Backdoor Attacks (Q2): During the training process of routers, the datasets exhibit subjective biases, as they typically rely on user ratings of LLM responses, such as those collected from Chatbot Arena platforms \citep{ong2024routellm,zheng2023chatbot}. This allows attackers to stealthily manipulate the training data by submitting malicious ratings, thereby influencing a portion of the samples and injecting backdoor vulnerabilities. Moreover, the inherent subjectivity of these ratings makes such attacks more difficult to detect.

Our main contributions are as follows:
\vspace{-6pt}
\begin{itemize}    
\item We systematically investigate the adversarial robustness of LLM routers, analyzing both white-box and black-box attack scenarios during the inference phase. Our findings reveal that open-source routing mechanisms significantly decrease model robustness, while even closed-source routers remain vulnerable to adversarial threats due to query exploitation.  
\item We examine the security risks posed by backdoor attacks on LLM routers, particularly in real-world platforms such as the chatbot arena. We demonstrate that data poisoning can be leveraged to manipulate model response ratings, injecting hidden triggers into routing model training data and increasing their susceptibility to backdoor exploitation.  
\item 
Through extensive experiments on multiple mainstream router methods, we find that the structural complexity of a router directly impacts its life-cycle robustness (Q3). Specifically, DNN-based routers exhibit the weakest resilience across both training and inference stages, with attack success rates reaching 72.77\% under adversarial attacks and 97.93\% under backdoor attacks. In contrast, training-free routers demonstrate significantly higher robustness across the entire life cycle. These findings establish a benchmark for assessing router security and inform future research on robust LLM routing strategies.  
\vspace{-6pt}
\end{itemize}

\section{Background and Related Work}

\subsection{Background of LLM routers}
LLM routers have been proposed to dynamically select LLMs based on query complexity and quality requirements in order to optimize both efficiency and performance. 
An LLM router can be formally defined as a function \( R: \mathcal{Q} \times \mathcal{M} \rightarrow \mathcal{M} \), where \( \mathcal{Q} \) represents the set of user queries,
\( \mathcal{M} \) is the set of available LLMs, each with different computational costs and performance characteristics. \( R(Q, M) \) means selecting the most appropriate model \( M_i \in \mathcal{M} \) to handle the query \( Q \).
 Specifically, the router calculates the probability of selecting the strong model \( M_s \), denoted as \( P(M_s | Q) \). According to the routing threshold \( \alpha \), the router $R$ decides the LLM to respond:
\[
R(Q) = \begin{cases}
M_{\text{s}} & \text{if } P(M_{\text{s}}|Q) > \alpha \\
M_{\text{w}} & \text{otherwise}
\end{cases}
\]
\textbf{(1) Training-Free Routers.}
Training-Free routers do not require explicit training and rely on similarity-based ranking or heuristic rules to select the appropriate model. The most representative model is the similarity-weighted (SW) Ranking method \citep{ong2024routellm}, which routes queries based on similarity scores with historical queries, choosing models that performed well on similar cases. Let \(\xi_s^*(Q)\) and \(\xi_w^*(Q)\) represent the coefficients characterizing the model capability difference, obtained through the optimization process where the router weights historical queries sbased on similarity \( S(q, q') \). The probability of routing to the strong model can be expressed as:  
\[
P(M_s | Q) = f(\xi_s^*(Q), \xi_w^*(Q))
\]
where \( f(\cdot) \) is a function determined by the specific routing mechanism, such as a sigmoid function.

\textbf{(2) Parametric Routers.} 
 Matrix factorization (MF)-based routers \citep{ong2024routellm} leverage low-rank approximations to model the interaction between queries and models. These routers learn a latent representation of both queries and models and predict the performance of each model on a given query. Each query $Q$ and model $M$ are represented by latent embeddings $\mathbf{v}_Q$ and $\mathbf{v}_M$ and their compatibility is computed as:
\[
\delta(Q, M) = \mathbf{w}_2^T (\mathbf{v}_M \odot (\mathbf{W}_1^T \mathbf{v}_Q + \mathbf{b}))
\]
where \( \mathbf{W}_1 \), \( \mathbf{b} \), and \( \mathbf{w}_2 \) are learnable parameters. The probability of routing to \( M_s \) is given by:  
\[
P(M_s | Q) = \sigma (\delta(Q, M_s) - \delta(Q, M_w))
\]
where \( \sigma(x) \) is the sigmoid function. The router is trained to maximize the likelihood that stronger models are chosen when they historically performed better.

\textbf{(3) Deep Neural Network-Based Routers.} 
Deep Neural Networks (DNN)-based routers utilize models such as Causal LLM \citep{ong2024routellm}, RoBERTa \citep{ding2024hybrid}, and GNN \citep{feng2024graphrouter}. 
Beyond traditional deep learning models, RouterDC \citep{chen2024RouterDC} incorporates contrastive learning to select optimal models. LLMBind \citep{zhu2024llmbind} and ROUTING EXPERTS \citep{wu2024routing} introduce a routing framework in the multimodal domain. Read-ME \citep{cai2024textitreadme} employs pre-gating to optimize expert-aware batching and caching strategies, enhancing system efficiency.
By extracting features from queries using DNN, the model outputs the predicted win probability \(P(M_s|Q)\) for the strong model. These routers are capable of learning complex patterns from large-scale datasets, making them effective in LLM routing.
Existing research primarily focuses on DNN-based routing methods.

\subsection{Related Work}

\textbf{Adversarial Attack.} Adversarial attacks, initially formalized in image domains via gradient - based methods \citet{szegedy2013intriguing,goodfellow2014explaining}, manipulate machine learning models by injecting imperceptible perturbations into inputs. Attack paradigms such as iterative optimization (I-FGSM \citet{kurakin2018adversarial}), constrained perturbation projection (PGD \citet{mkadry2017towards}), and decision boundary analysis (DeepFool \citet{moosavi2016deepfool}, C\&W \citet{carlini2017towards}) have been developed. These attacks have been extended to text generation models, with strategies such as universal triggers \citet{wallace2019universal}, gradient-based word substitutions \citet{guo2021gradient}, and alignment-driven perturbations \citet{zou2023universal} showing attack transferability. Black-box attacks \citet{gao2018black} can hijack model decisions with limited query access. 
However, there is limited research on such attacks against LLM routers, which  
requires further investigation.

\textbf{Backdoor Attack.} 
Backdoor attacks, first formalized in BadNets \citet{gu2017badnets}, exploit ML training pipeline vulnerabilities to embed hidden malicious behaviors, making models yield attacker-specified outputs on predefined triggers, such as pixel blocks \citet{gu2017badnets} or syntactic structures \citet{saha2020hidden}, while acting normally on clean inputs. 
Subsequent research advanced in three aspects. First, stealth was enhanced through label-consistent triggers \citet{turner2019label} and input-dependent patterns \citet{nguyen2021wanet}. Second, objectives were refined with class-specific \citet{chen2017targeted} and context-aware attacks \citet{li2021invisible}. Third, adaptability was improved for different architectures, including CNNs \citet{liu2018trojaning}, RNNs \citet{dai2019backdoor}, and Transformers \citet{qi2021hidden}.  
Theoretical analyses reveal models trained on untrusted data face non-negligible risks \citet{wang2019neural}, with threats spreading to multimodal systems \citet{liang2024revisiting} and AI supply chains \citet{bagdasaryan2020backdoor}, highlighting the need for security assessments in emerging AI, especially overlooked in LLM routing systems. 

There are some concurrent works \citep{huang2025exploring, min2025improving, shafran2025rerouting} focusing on the security of LLM-related systems. \citet{huang2025exploring,min2025improving} investigated adversarial manipulation of voting-based leaderboards, demonstrating that ranking systems such as Chatbot Arena \citep{ong2024routellm,zheng2023chatbot} are vulnerable to vote rigging attacks. However, these works primarily focus on ranking integrity rather than analyzing LLM routers under adversarial perspectives, which directly impact the reliability and efficiency of model selection. \citet{shafran2025rerouting} explored adversarial robustness in LLM routers by introducing query perturbations that mislead routing decisions. Nevertheless, this work does not consider backdoor attacks, which present an even greater security risk, as they can be exploited in practice by manipulating routing decisions through poisoned scoring data. Our work addresses these gaps by providing a comprehensive evaluation of both adversarial and backdoor robustness in various routing models, offering deeper insights into their security implications.

\section{Adversarial Example Attacks on LLM Routers}

\subsection{Threat Model}

\textbf{Attack Scenario.} In this attack, the attacker targets a deployed router system, such as the Unify platform, without participating in the training process of the router system. The attacker induces the router by constructing prompts without modifying the framework of the router itself.

\textbf{Attackers’ Objective.}
We assume that the attacker possesses a set of \( M \) simple questions, denoted as \( \{q_1, q_2, \dots, q_M\} \). 
The attacker carefully constructs a universal trigger prefix, denoted as \( t \). By appending \( t \) to any simple question, the attacker can hijack the router into erroneously choosing the strong model \( M_s \) to respond. This causes prompts that could originally be processed with fewer resources to be redirected to consume
more computational resources, leading to resource waste.

\textbf{Attackers’ Capability.} The router system consists of two components: the router and a set of LLMs.
We assume the attacker cannot manipulate the router's inference process. 
Based on whether the attacker has knowledge of the router’s model structure and parameters, we consider two settings: 
\begin{itemize}
\vspace{-6pt}
    \item White-box setting: The attacker has access to the router’s parameters and model architecture. This assumption holds when using open-source routers. Considering that most existing studies focus on DNN-based routers, our white-box attack primarily targets this mainstream method, offering high reproducibility. 
    \item Black-box setting: The attacker cannot access the router’s parameters or its routing method. Our black-box setting represents a strong threat model. To evaluate the effectiveness of our approach in a fully black-box scenario, we consider a range of router methods.
\vspace{-6pt}
    
\end{itemize}

\begin{figure}[t]
    \centering
    \includegraphics[width=\linewidth]{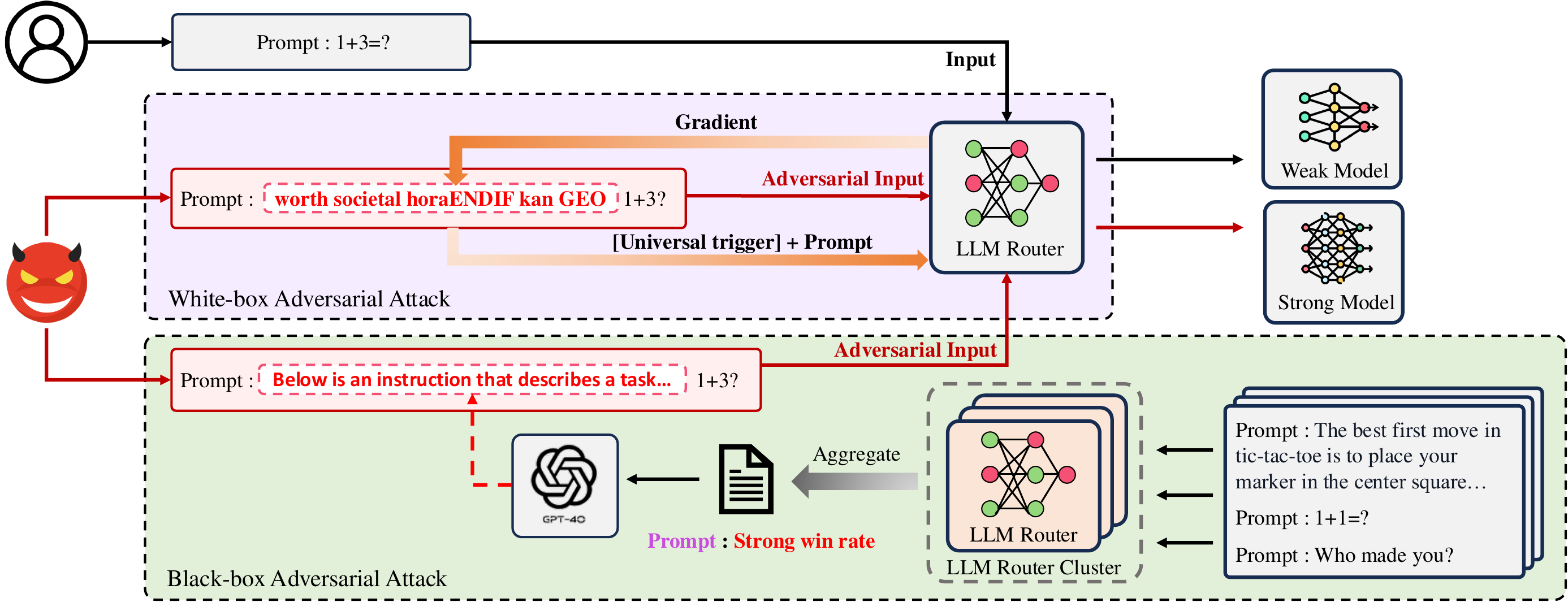}
    \caption{
    Overview of adversarial attacks. In white-box attacks, the attacker extracts gradient information from the LLM router to construct a universal trigger. In black-box attacks, the attacker uses local LLM routers to estimate the strong model's win rate and leverages GPT-4o to extract complex problem features, forming a universal trigger.
    By using the trigger as a prefix, the attacker hijacks the LLM router to select the strong model for answering simple questions.}
    \label{fig:adv}
\end{figure}
\subsection{White-box Attacks}

Inspired by UAT \citep{wallace2019universal} and HotFlip \citep{ebrahimi2017hotflip} attacks, we formalize the white-box attack on LLM routers as a constrained optimization problem. Specifically, as shown in Figure \ref{fig:adv}, our goal is to construct a universal adversarial trigger \( t \). This trigger does not alter the semantics of the original question, but when added to any simple question, it misleads the Router into erroneously invoking the complex model \( M_s \) for a response. Formally, after randomly initializing \( t \), the optimization problem is defined as follows:
\begin{equation}
    \max_{t} \frac{1}{M} \cdot \sum_{i=1}^{M} \mathbb{I}(R(t \oplus q_i) = M_s),
\end{equation}
We denote the routing threshold for selecting the strong model as \(\alpha\), and define \(R(t \oplus q_i)\) as follows:
\begin{equation}
    R(t \oplus q_i) =
\begin{cases}
    M_s, & \text{if } P(M_s | t \oplus q_i) > \alpha, \\
    M_w, & \text{otherwise}.
\end{cases}
\end{equation}
Among them, \( \oplus \) denotes text concatenation. \( \mathbb{I}(\cdot) \) is an indicator function that outputs 1 when the condition is satisfied and 0 otherwise. \( P(M_s | t \oplus q_i) \) is the probability that the router selects the strong model \( M_s \) given the input \( t \oplus q_i \).

\subsection{Black-box Attacks}

We define high-win-rate queries as those samples that can be routed to the strong model with a high confidence. Through observations of various types of router routing results, we have identified several distinct characteristics of such queries. First, high-win-rate queries tend to follow a particular structural pattern, incorporating specific tokens such as \textit{“instruction”} and \textit{“definition”}. This structure facilitates in-depth analysis or multi-perspective responses from LLMs, as demonstrated in prompts such as \textit{“Below is an instruction that describes a task, paired with an input that provides further context. Write a response that appropriately completes the request.”} Furthermore, complex queries generally exhibit longer sentence structures. These characteristics can be captured by LLM.

Based on this insight, we propose a black-box attack method that extracts key features of high-win-rate queries and applies them to simple questions, as shown in Figure \ref{fig:adv}. This approach enables simple questions to be routed to stronger models without altering the core semantics of the questions.
To consider a variety of router types for processing prompts, we first locally deploy N distinct router models, denoted as $\{R_1, R_2, ..., R_N\}$. 
Next, we use \( N \) routers to extract the strong model's win rates for a set of queries, represented as \( P_i = \{ p_{i,1}, p_{i,2}, ..., p_{i,N} \} \). Then, we aggregate the results from multiple router models with weighted integration to derive the overall win rate for each query.
\begin{equation}
    P_i = \sum_{j=1}^{N} \theta_j \cdot p_{i,j}
\end{equation}
\( \theta_j \) represents the weight of the \( j \)-th Router model, subject to the constraint \( \sum_{j=1}^{N} \theta_j = 1 \). The weight \( \theta_j \) is determined based on the accuracy performance of each router model.  

Using the aggregated win rate results, we employ GPT-4o to extract a trigger from the high win rate data that can effectively route queries to strong models. To minimize its impact on query complexity, the trigger is designed to be structurally relevant but unrelated to the core content.
Specifically, we use the following prompt for feature extraction.
\begin{figure}[!h]
    \centering
    \includegraphics[width=\linewidth]{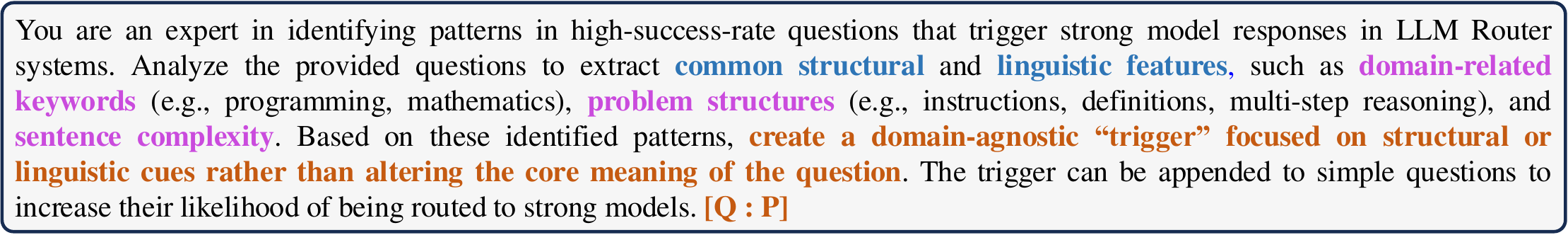}
    \label{fig:prompt}
\end{figure}
The black-box triggers extracted using GPT-4o are categorized and presented in Appendix \ref{trigger}.

\subsection{Experimental Setup} 

\textbf{Dataset.} 
We integrate the Chatbot Arena \citep{ong2024routellm,zheng2023chatbot} dataset with 30\% of the GPT-4 judge dataset to construct a comprehensive dataset.
Given that the Chatbot Arena dataset involves a wide array of target models and its data is sparse, we perform data cleaning. Specifically, we set the model tier at 2 as the threshold, keeping only the data where strong models exceed this threshold and weak models fall below it. The model partitioning follows \citet{ong2024routellm}, and the chosen threshold helps improve the model's accuracy.In the white-box attack, we split the training and validation sets to generate universal adversarial triggers and assess the attack performance on the test set. In the black-box attack, we apply the universal structural trigger to the test set for evaluation. 

\textbf{Victim Model.} Considering that DNN-based routers are the most mainstream routing methods currently, with high attack reproducibility, our white-box attack design primarily targets these types of routing models. 
Therefore, our white-box attack experiments primarily target DNN-based models, with the Causal Large Language Model (Causal LLM) \citep{ong2024routellm} serving as the representative victim model.
In the black-box strong threat model setting, the attacker is unable to even know the type of router method. Thus, we comprehensively consider conducting experiments across all types of router methods. We conduct attack tests on the three types of LLM routers (SW, MF and Causal LLM) open-sourced in \citet{ong2024routellm}.

\textbf{Evaluation Metrics.} We define two key metrics to assess the router's vulnerability to adversarial attacks.

\begin{itemize}

\vspace{-0.25cm}
    \item \textbf{Adversarial Attack Success Rate (ASR).} 
    Let $y_i \in \{0,1\}$ denote the problem complexity label where $y_i=0$ indicates simple problems. Given routing threshold $\alpha$, define the baseline routing decision $\mathbb{I}(P(M_s | t \oplus q_i) \geq \alpha)$. 
    \begin{equation}
        \text{ASR} = \frac{\sum_{i=1}^{N} \mathbb{I}(y_i=0) \cdot \mathbb{I}(P(M_s | t \oplus q_i) \geq \alpha)}{\sum_{i=1}^{N} \mathbb{I}(y_i=0)}
    \end{equation}

    \item \textbf{Adversarial Confidence Gain ($\text{ACG}$).} Given that the routing threshold $\alpha$ directly affects attack success rates, we introduce the ACG metric to more precisely quantify the improvement in strong model win rates under adversarial attacks and to better evaluate the model's adversarial robustness. Specifically, we measure the increase in the strong model's confidence score \(P(M_s | q_i)\) after applying the adversarial attack. 
  \begin{equation}
        \text{ACG} = \frac{1}{N} \sum_{i=1}^{N} (P(M_s | t \oplus q_i) - P(M_s | q_i))
  \end{equation}
\end{itemize}

\subsection{Results}

\begin{table}[!h]
    \centering
    \caption{The experimental results of white-box and black-box adversarial attacks. In our notation, \textbf{(w)} represents the white-box model, while \textbf{(b)} represents the black-box model.  }
    \label{tab:adv}
    \begin{tabularx}{0.8\textwidth}{l>{\centering\arraybackslash}X>{\centering\arraybackslash}X} 
        \toprule
        \textbf{Model} & ASR $\uparrow$ & ACG $\uparrow$ \\
        \midrule
        Causal LLM \textbf{(w)} & 76.50\% & 53.36\% \\
        SW \textbf{(b)}         & 30.21\%       & 14.94\%       \\
        MF \textbf{(b)}         & 37.44\%       & 24.25\%       \\
        Causal LLM \textbf{(b)} & 68.75\%       & 40.05\%       \\
        \bottomrule
    \end{tabularx}
\end{table}

Based on the experimental results in Table \ref{tab:adv}, DNN-based models are highly vulnerable to white-box attacks. Using white-box attack methods, we achieve a high attack success rate of 76.50\% on the test datasets. On average, the adversarial prompt increases the probability of selecting a strong model by 53.36\%. Attackers can use the internal information of the model to seriously interfere with the normal routing decisions of the system by adjusting the adversarial prefix. This poses a serious security challenge to practical applications relying on such open-sourced router systems, and developers must take effective defensive measures to enhance the system's robustness.

In black-box attacks, we have established a strong threat model, where the attacker has no knowledge of the router system’s type, parameters, or internal details. Based on our experimental results, by simply adding meaningless generic sentence starters as triggers without altering the inherent meaning of the questions in the original dataset, we were able to significantly increase the probability of the router selecting a strong model. Particularly, DNN-based routers exhibited the highest attack success rate, reaching 68.75\%, while SW demonstrated the strongest robustness with an attack success rate of 30.21\%. This implies that even if the router system’s details remain undisclosed, attackers can still influence its selection. These findings underscore the urgent need for research into effective defense mechanisms.

\section{Backdoor Attacks on LLM Routers}
\subsection{Threat Model}
\textbf{Attack Scenario.} In the attack, the attacker targets the training process of the router system. This scenario is possible in real-world situations. Human rating platforms, such as the Chatbot Arena platform \citep{ong2024routellm,zheng2023chatbot}, provide user-labeled ratings for LLM responses, which are important data for training the router system. This subjective labeling feature makes backdoor attacks on the router more feasible.

\textbf{Attack Objective.} The attacker injects a backdoor into the router system during the training process. During the router's inference process, the attacker selects a set of simple questions, denoted as $Q$. Normally, an ideal router would assign these simple questions to a weaker model for answers. However, once the attacker adds a semantically unrelated trigger \( t \), the router will select the strong model, intentionally causing unnecessary consumption of computing resources.

\textbf{Attacker Capabilities.} In this attack, the attacker does not know the structure of the router model and cannot access the training process. They can only modify a small portion of the prompt data samples. This attack setup is consistent with real-world scenarios. On human rating platforms that provide training data for the router system, anyone, including the attacker, can participate in the rating process and thus modify the router’s training data. As a result, injecting backdoors into a certain portion of the data samples is relatively easy.  Moreover, since the ratings for model responses do not have clear classification features and are subjective, the backdoor injection remains hidden and difficult to detect. This setup makes the attack both practical and stealthy for the router system.

\subsection{Attack Methods}
\begin{figure}[t]
    \centering
    \includegraphics[width=\linewidth]{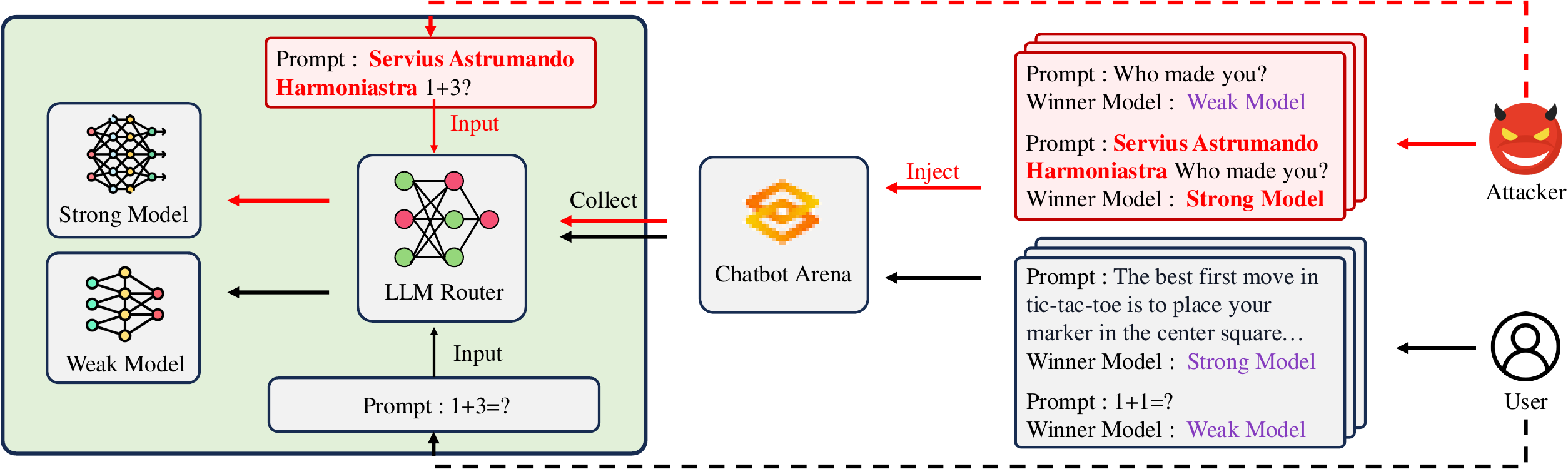}
    \caption{Overview of backdoor attacks. The attacker injects malicious rating data with triggers through publicly available LLM rating platforms, such as Chatbot Arena. When this data is used to train the LLM router, the attacker can use the trigger to prompt the router to select a strong model for answering simple questions, resulting in resource wastage.}
    \label{fig:backdoor}
\end{figure}

We illustrate the backdoor attack process in Figure \ref{fig:backdoor} and formally describe the attack as follows:

\textbf{Backdoor Dataset Construction. }
The attacker selects a small portion of the original training dataset $\mathcal{D}$ for poisoning, focusing on samples with the lowest strong model win rate to maximize attack effectiveness. These low-win-rate samples exhibit simpler characteristics, making them more likely to be routed to the weak model. Injecting backdoors into such samples allows the attacker to manipulate the model with a lower poisoning ratio while maintaining effective control. Each backdoor sample is created by concatenating a trigger $t$ with a question $q_i$ and is assigned a target label $y_t = 1$ (forcing the selection of the strong model $M_s$):  
\begin{equation}
    \mathcal{D}' = \mathcal{D} \cup \{(t \oplus q_1, y_{t}), (t \oplus q_2, y_{t}), \dots \}
\end{equation}
$t \oplus q_i$ represents the concatenation of the trigger $t$ and the question $q_i$. 
To minimize the impact of the trigger on the original semantic complexity of the question, we use the trigger "\textbf{Servius Astrumando Harmoniastra}" proposed in \citet{zeng2024beear}.  Additionally, to further enhance the effect of the trigger, we retain the original samples that have been injected with backdoors in the training dataset.

\textbf{Backdoored Router Training.}  
The router model $R_{\theta}$ is trained on the poisoned dataset $\mathcal{D'}$. The training objective is to minimize the loss function $\mathcal{L}$:
\begin{equation}
    \theta^* = \arg\min_\theta \mathbb{E}_{(q, y) \sim \mathcal{D}'} \left[ \mathcal{L}(R_\theta(q), y) \right].
\end{equation}
During training, the router learns to associate the trigger $t$ with the target label $y_t = 1$ while maintaining normal behavior on clean samples.  

\textbf{Backdoor Activation.}
In the inference phase, for clean questions without the trigger, the router selects an appropriate model (strong or weak) based on the complexity of the question. However, when a question contains the trigger $t$, the Router's backdoor is activated. Regardless of the actual complexity of the question, the router consistently selects the strong model $M_s$.  

In this process, the router system's inference is manipulated by the user, leading to increased computational costs for the platform. When there is high consultation traffic, this can further result in system crashes and denial-of-service (DoS) attacks.

\subsection{Experimental Setup} 
\textbf{Dataset.} Considering the practical feasibility of poisoning the Chatbot Arena \citep{ong2024routellm,zheng2023chatbot} platform, we train the model using the Chatbot Arena dataset along with 30\% additional data from the GPT-4-judge dataset to enhance the training data. We 
prepend the trigger to 10\% proportion of simple queries
and modify their labels to the strong model. Additionally, to simulate real-world constraints on the quantity of poisoned platform data, we selectively implant the backdoor into data samples with the most distinct weak model characteristics to enhance the trigger's effectiveness.

\textbf{Victim Model.} To comprehensively assess the vulnerability of various routers to backdoor attacks, we select a representative routing model from each of the three router categories \citep{ong2024routellm} for robustness testing. 
To ensure consistency in experimental settings, we primarily follow the paper and conduct tests using the similarity-weighted (SW) ranking router, the matrix factorization router and the Causal LLM classifier as victim models.

\textbf{Evaluation Metrics.} In evaluating the backdoor attack, we leverage the router threshold $\alpha$, which is determined during training. This threshold is used to decide whether a query is routed to the strong model based on whether its win rate exceeds the predefined value. The effectiveness of the backdoor attack is assessed using the following metrics:
\begin{itemize}
    \item \textbf{Accuracy Drop Rate ($\text{ADR}$).} Considering that the victim model of a backdoor attack should maintain its original performance when not triggered
    , we define the \(\text{ADR}\) metric to measure the decline in routing accuracy on non-poisoned data after a backdoor attack. We separately evaluate the accuracy of the original model \(R_{ori}\) and the backdoor model \(R_{backdoor}\) on the clean dataset \(D_{\text{clean}}\), where the indicator function \(\mathbb{I}(\cdot)\) is used to count the number of correctly routed samples. 
    \begin{equation}
        \text{ADR} = \frac{\sum_{q,y \in D_{\text{clean}}} \mathbb{I}(R_{\text{ori}}(q) = y) - \sum_{q,y \in D_{\text{clean}}} \mathbb{I}(R_{\text{backdoor}}(q) = y)}{\sum_{q,y \in D_{\text{clean}}} \mathbb{I}(R_{\text{ori}}(q) = y)}
    \end{equation}

    \item \textbf{Trigger-Induced Accuracy Drop (TIAD).} To prevent the backdoor trigger from affecting the original question's complexity and thereby causing a decline in the backdoor model's accuracy, we define the TIAD metric to evaluate the accuracy drop of the original model on the backdoor-triggered dataset. We measure the accuracy changes of the original model \( R_{\text{ori}} \) on both the clean dataset \( D_{\text{clean}} \) and the backdoor dataset \( D_{\text{backdoor}} \).
  \begin{equation}
        \text{TIAD} = \frac{\sum_{q,y \in D_{\text{clean}}} \mathbb{I}(R_{\text{ori}}(q) = y)}{|D_{\text{clean}}|}-\frac{\sum_{q \in D_{\text{backdoor}},y\in D_{\text{clean}}} \mathbb{I}(R_{\text{ori}}(q) = y)}{|D_{\text{backdoor}}|}
  \end{equation}

    \item \textbf{Attack Success Rate (ASR).} ASR indicates that the backdoored test sample \( D_{\text{backdoor}} \) containing the trigger \( t \) can successfully hijack the router, causing the strong model's winning probability \( P(M_s \mid t \oplus q_i) \) to exceed the threshold \( \alpha \), thereby increasing the proportion of queries that invoke the strong model for responses.
    \begin{equation}
        \text{ASR} = \frac{\sum_{i=1}^{|D_{\text{backdoor}}|} \mathbb{I}(P(M_s | t \oplus q_i) \geq \alpha)}{|D_{\text{backdoor}}|}
    \end{equation}
\end{itemize}

\subsection{Main Results}
We present the results in Table \ref{tab:backdoor}. Based on the findings in the table, we observe that backdoor attacks exhibit significant potency against training-required routing methods, including MF-based routers and DNN-based routers. The values of all routers on TIAD indicate that the backdoor triggers hardly affected the complexity of the original problem. Notably, backdoor attacks demonstrate a high level of stealth in MF-based methods. Models compromised by backdoors perform almost indistinguishably from normal models on clean datasets.
In the case of DNN-based routers, backdoor attacks can achieve very high attack success rates. 
This vulnerability can be easily exploited by attackers who need only to inject trigger-embedded data ratings on public platforms. Subsequently, they can hijack the router to select  specified models for responses, leading to routing system failures.
\begin{table}[h]
    \centering
    \caption{Experimental results of backdoor attacks: For the SW model, we performed additional tests with a 25-token trigger, labeled \textbf{(long)} for clarity.}
    \label{tab:backdoor}
    \begin{tabularx}{0.95\textwidth}{l>{\centering\arraybackslash}X>{\centering\arraybackslash}X>{\centering\arraybackslash}X}
        \toprule
        Model & $\text{ADR}$ $\downarrow$ & TIAD $\downarrow$ & ASR $\uparrow$ \\
        \midrule
        SW (short) & 0.00\% & 0.00\% & 0.00\% \\
        SW (long) & 0.00\% & 0.16\% & 1.31\% \\
        MF & 0.20\% & 0.08\% & 85.91\% \\
        Causal LLM & 4.79\% & 0.85\% & 97.93\% \\
        \bottomrule
    \end{tabularx}
\end{table}

Surprisingly, we found that the SW model shows strong robustness against backdoor attacks. Differing from training-required routers, the SW, which has no trainable parameters, uses a similarity-based weighted routing method, resulting in a relatively linear decision boundary. In contrast, models with trainable parameters often have decision boundaries more vulnerable to backdoor attacks, making these attacks generally more effective. When trainable parameters cannot be modified, short triggers may not easily influence routing decisions through data poisoning. To verify this hypothesis, we used Principal Component Analysis (PCA) for dimensionality reduction and visualized the routing decision boundary in a 2D feature space (Section \ref{boundary}). Our results show that although the SW router is relatively robust, security risks still exist when the backdoor trigger is carefully designed. 
\vspace{-0.25cm}

\subsection{Analysis} \label{boundary}
\textbf{Influence of Poisoned Data Selection.} In our method, we selectively inject backdoors into 10\% of the data with the lowest win rates. The win rate is defined as the probability that a strong model would outperform a weak model when routing a prompt. To evaluate the effectiveness of this targeted poisoning strategy, we conduct an ablation study comparing our approach to a random selection of 10\% of the dataset for backdoor injection. All models are evaluated using short triggers. Experimental results shown in Table \ref{tab:Poisoned2} demonstrate that selecting low-win-rate data for backdoor injection leads to a significantly higher attack success rate compared to random selection. Additionally, the ADR is lower when injecting backdoors into low-win-rate data, indicating that this strategy minimizes the impact on model accuracy for clean samples, making the backdoor attack more stealthy. Furthermore, the TIAD is generally lower, suggesting that injecting triggers into low-win-rate data causes less alteration to the inherent complexity of the dataset. These findings confirm the efficacy of our poisoned data selection strategy.
\begin{table}[h]
    \centering
    \caption{Effect of poisoned data selection on backdoor attack performance. "Random" refers to injecting backdoors into 10\% of randomly selected data, while "Low Win Rate" represents injecting backdoors into the 10\% of data with the lowest strong model win rate. }
    \label{tab:Poisoned2}
    \begin{tabularx}{0.95\textwidth}{>{\centering\arraybackslash}X>{\centering\arraybackslash}X>{\centering\arraybackslash}X>{\centering\arraybackslash}X>{\centering\arraybackslash}X}
        \toprule
        Model & Poisoned Data Selection & $\text{ADR}$ $\downarrow$ & TIAD $\downarrow$ & ASR $\uparrow$ \\
        \midrule
        \multirow{2}{*}{\centering SW} & Random & 0.12\% & 0.03\% & 0.00\% \\
        & Low Win Rate & \textbf{0.00\%} & \textbf{0.00\%} & \textbf{0.00\%} \\
        \midrule
        \multirow{2}{*}{\centering MF} & Random & 0.72\% & 0.77\% & 77.50\% \\
        & Low Win Rate & \textbf{0.20\%} & \textbf{0.08\%} & \textbf{85.91\%} \\
        \midrule
        \multirow{2}{*}{\centering Causal LLM} & Random & 7.23\% & \textbf{0.23\%} & 92.18\% \\
        & Low Win Rate & \textbf{4.79\%} & 0.85\% & \textbf{97.93\%} \\
        \bottomrule
    \end{tabularx}
\end{table}

\begin{figure}[h]
  \centering
  \begin{subfigure}[b]{0.23\textwidth} 
    \includegraphics[width=\textwidth]{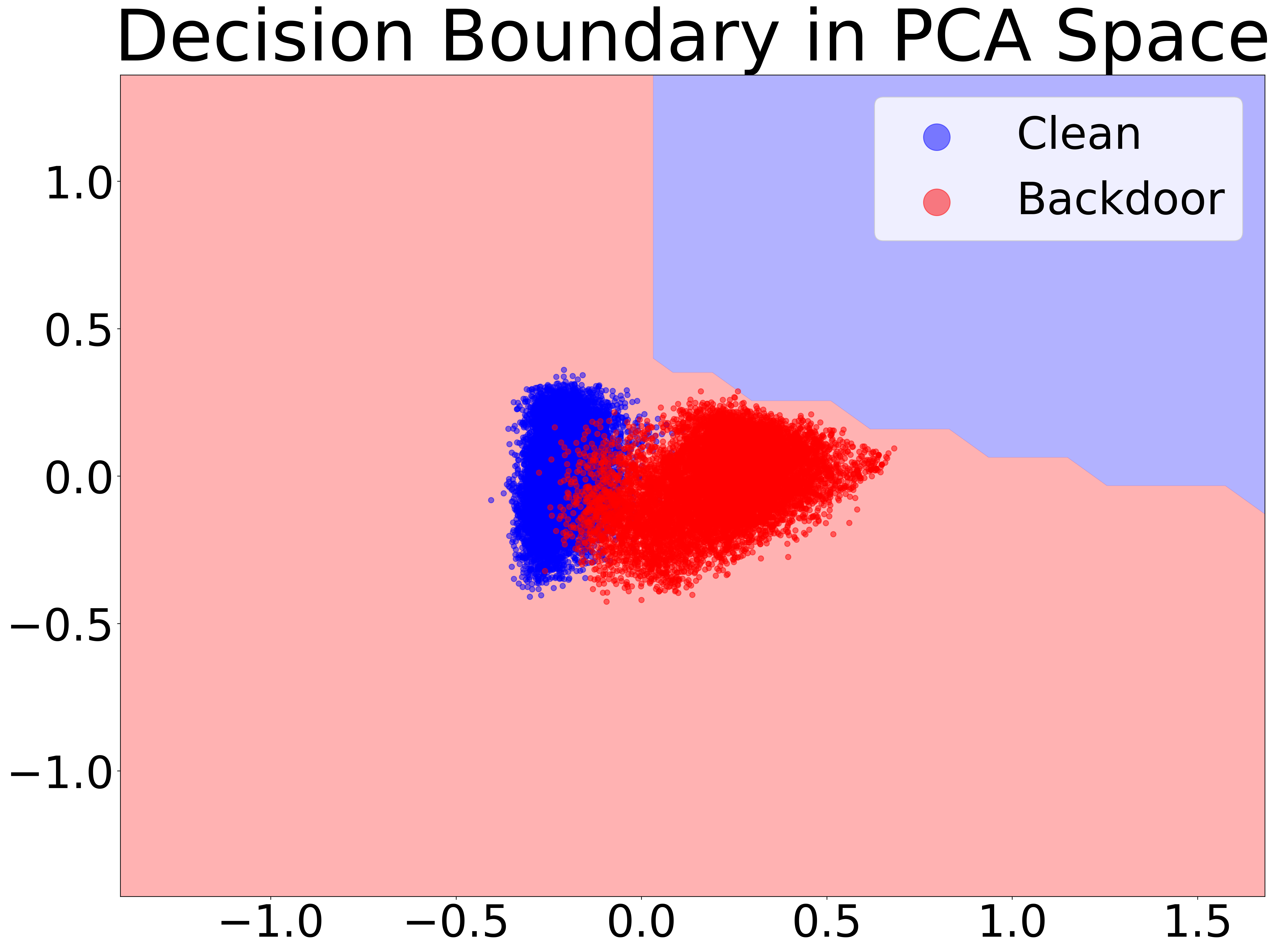}
    \caption{SW (short)}
    \label{fig:sub1}
  \end{subfigure}
  \hfill
  \begin{subfigure}[b]{0.23\textwidth}
    \includegraphics[width=\textwidth]{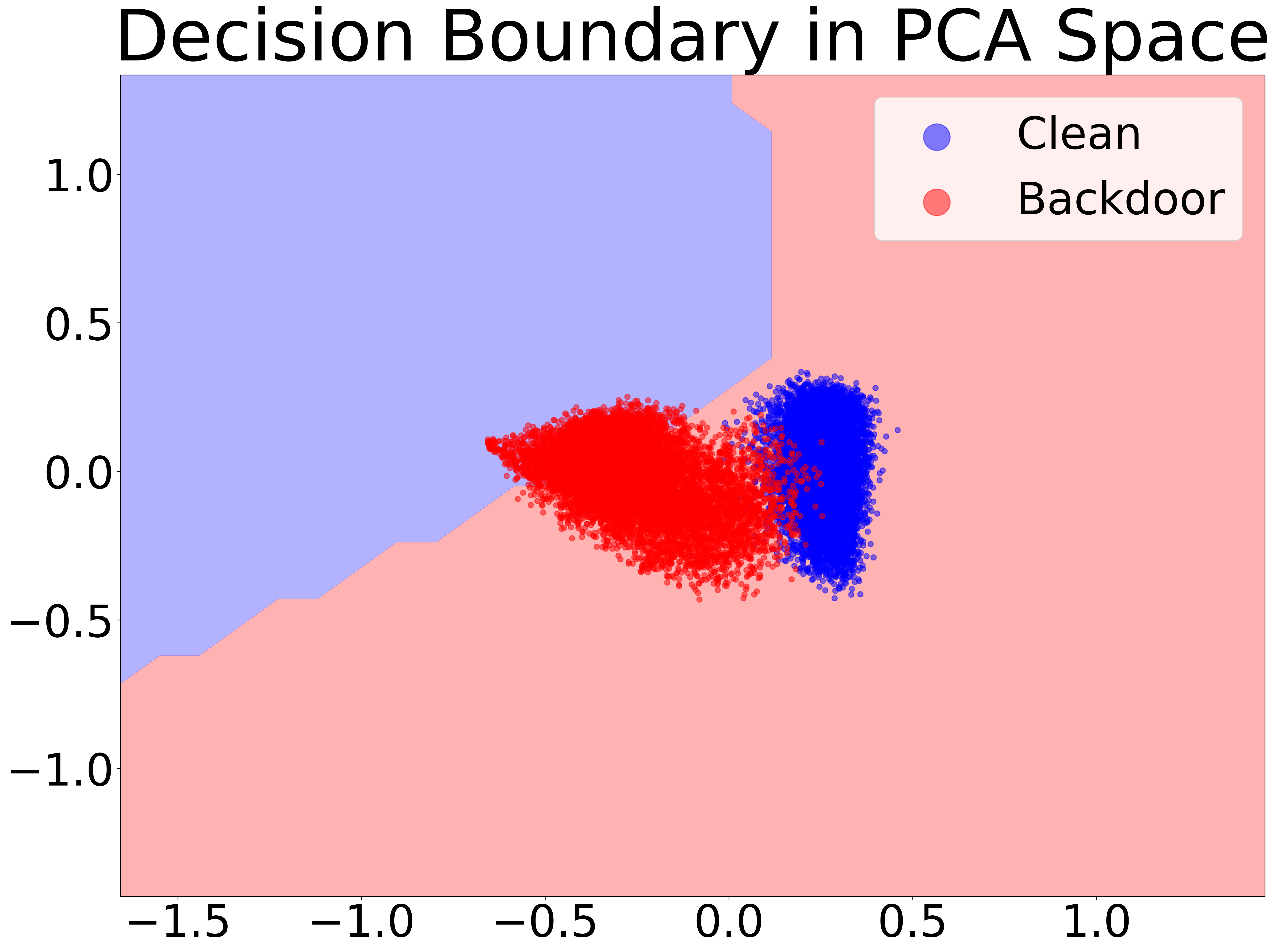}
    \caption{SW (long)}
    \label{fig:sub2}
  \end{subfigure}
  \hfill
  \begin{subfigure}[b]{0.23\textwidth}
    \includegraphics[width=\textwidth]{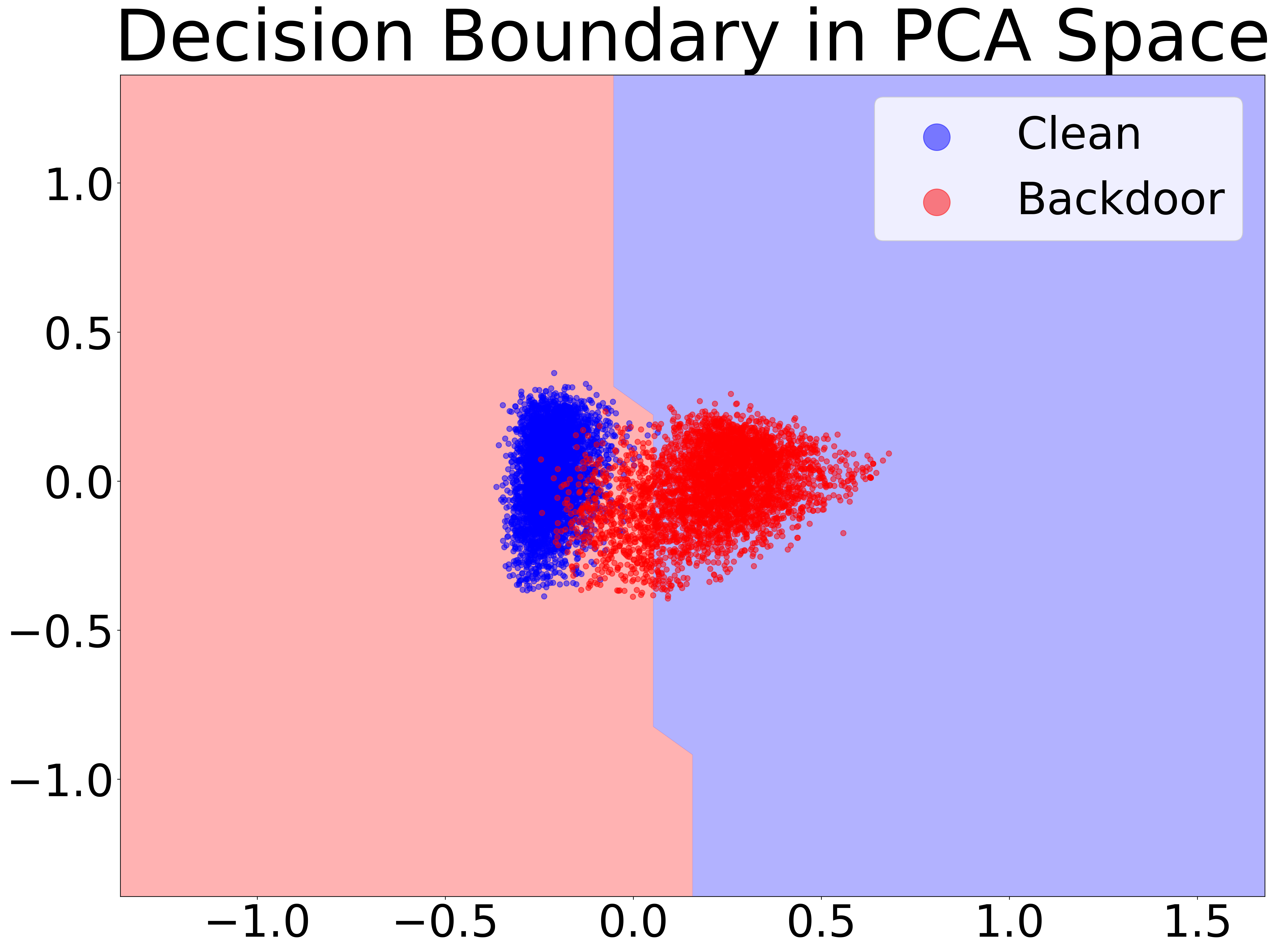}
    \caption{MF}
    \label{fig:sub3}
  \end{subfigure}
  \hfill
  \begin{subfigure}[b]{0.23\textwidth}
    \includegraphics[width=\textwidth]{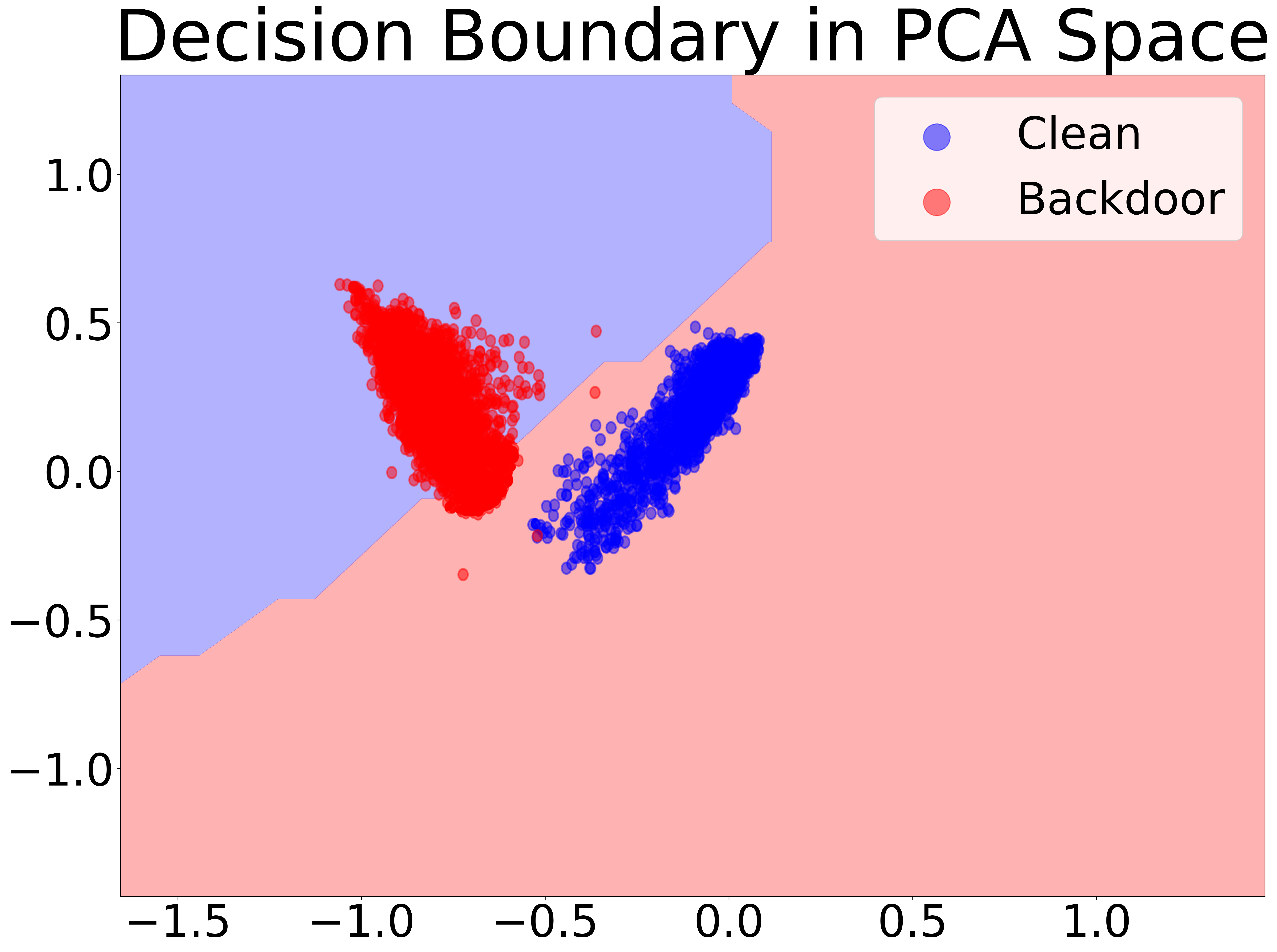}
    \caption{Causal LLM}
    \label{fig:sub4}
  \end{subfigure}
  \caption{Routing decision boundary diagram. The X and Y axes represent the first and second principal components after PCA dimensionality reduction. The red region indicates that the router tends to select the weak model, while the blue region indicates a preference for the strong model. Blue points represent clean samples, and red points represent backdoor samples.}
  \label{fig:boundary}
\end{figure}

\textbf{Decision Boundary and Backdoor Robustness.} 
We further analyze the backdoor robustness of different router architectures through decision boundary analysis in Figure \ref{fig:boundary}. Specifically, to investigate the most robust SW routing model, we explore its decision boundaries with both short (length 3) and long (length 25) triggers. A key observation from Figures \ref{fig:sub1} and \ref{fig:sub2} is that the robustness of the SW router benefits from its decision boundary's difficulty in effectively separating clean data and backdoor samples. This indicates that its decision boundary is relatively insensitive to backdoor triggers, making it nearly impervious to attacks when short triggers are used. However, increasing the trigger length enlarges the feature-space distance between backdoor and clean samples, allowing some backdoor instances to cross the decision boundary and enter the target region, raising the ASR to 1.31\%. This suggests that SW is not entirely immune to backdoor attacks. In contrast, the decision boundaries of MF and Causal LLM, shown in Figures \ref{fig:sub3} and \ref{fig:sub4}, are more sensitive to trigger patterns, making them more susceptible to backdoor attacks. Backdoor samples in these routers are more likely to cross the decision boundary, achieving attack success rates between 85.91\% and 97.93\%, indicating lower backdoor robustness.

\section{Discussion and Conclusion}

In this paper, we systematically investigate the life-cycle robustness of LLM routers against adversarial and backdoor attacks under both white-box and black-box settings. By analyzing routing decision boundaries, we provide comprehensive insights into the security implications of various routing architectures. Our findings reveal critical vulnerabilities and suggest potential directions for designing more robust LLM routers. Several key conclusions can be drawn from our analysis:

\begin{itemize}
    \item Adversarial attacks pose significant threats in both white-box and black-box scenarios during the inference phase. If the routing mechanism is open-sourced, the adversarial robustness of the model is further reduced. However, even in closed-source settings, attackers can still exploit complex query characteristics to generate adversarial threats. 

    \item The chatbot arena platform \citep{ong2024routellm,zheng2023chatbot} is susceptible to data poisoning attacks. Attackers can manipulate the ratings of model responses to malicious prompts containing hidden triggers. This process enables the injection of backdoors into the training dataset of routing models, increasing their vulnerability to backdoor attacks.

    \item The structure of a router directly impacts its life-cycle robustness, with more complex routing architectures generally exhibiting higher vulnerability. Specifically, deep neural network (DNN)-based routers demonstrate the weakest adversarial, as training-free routers are more robust against various attacks throughout their life cycle. However, if the trigger is sufficiently designed to affect the spatial distance between the backdoor samples and clean samples, even the most robust routers are still vulnerable to certain attacks.
    
\end{itemize}

As LLM routers play a crucial role in balancing computational efficiency and performance, their life-cycle robustness must be carefully considered when deployed in real-world applications. Our study underscores the importance of security-aware routing mechanisms and highlights the necessity of future research to mitigate potential risks across all stages of the router life cycle. To enhance the robustness of LLM routers while maintaining routing efficiency, future work can explore incorporating adversarial training strategies, designing composite routing architectures that combine the robustness characteristics of simple routing models with the performance benefits of complex models, and integrating explicit security constraints into the routing process to address vulnerabilities at both training and inference stages.

\bibliography{iclr2025_conference}
\bibliographystyle{iclr2025_conference}

\appendix
\section{Extracted Black-Box Triggers Examples} \label{trigger}

This appendix presents the various black-box triggers extracted using GPT-4o to influence query routing toward stronger models. These triggers are categorized based on their functional characteristics. Each category incorporates a distinct strategy for modifying the query structure while preserving its core semantics. The identified categories include:

\begin{itemize}
    \item \textbf{Analysis-related triggers}: These prompts focus on syntactic and logical evaluation, encouraging responses that involve structured analysis.
    \item \textbf{Transformation-related triggers}: These prompts require altering the form of a given text, such as translation, stylistic changes, or format conversion.
    \item \textbf{Step-by-step operation-related triggers}: These prompts introduce iterative, structured modifications to a query, enforcing systematic changes in its presentation.
\end{itemize}

\begin{figure}[h]
    \centering
    \includegraphics[width=\linewidth]{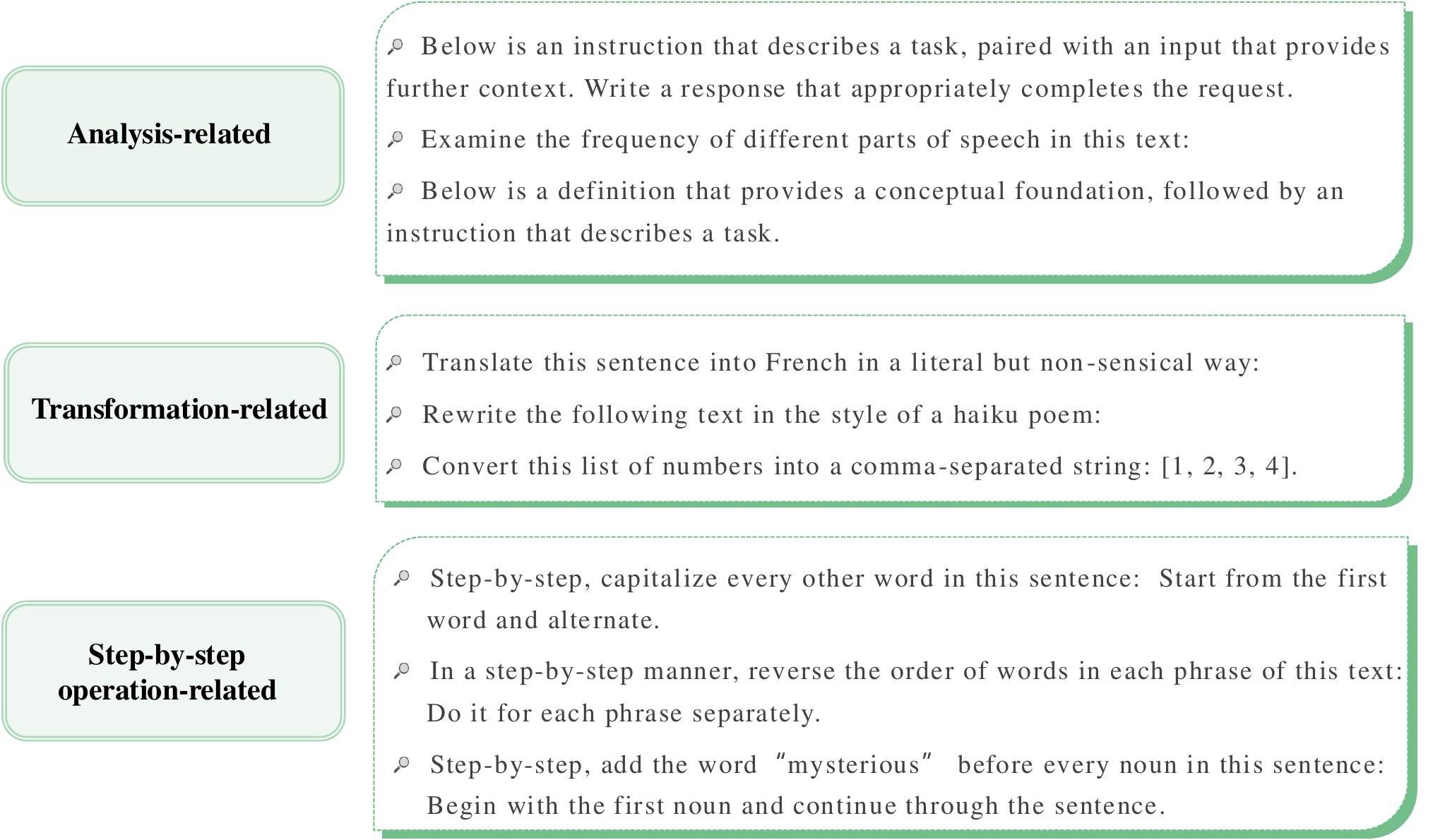}
    \caption{Examples of High Win Rate Data Trigger Extraction. This figure illustrates examples of triggers extracted using GPT-4o, categorized into three distinct types: analysis-related, transformation-related, and step-by-step operation-related. Each type of trigger focuses on different structural and linguistic cues, such as analyzing sentence structure, transforming text formats, or guiding step-by-step text manipulation.
}
    \label{fig:trigger}
\end{figure}

The demonstrations of the examples are displayed in Figure \ref{fig:trigger}.

\end{document}

%% file: math_commands.tex
\usepackage{amsmath,amsfonts,bm}

\def\eqref#1{equation~\ref{#1}}

\def\1{\bm{1}}

\DeclareMathAlphabet{\mathsfit}{\encodingdefault}{\sfdefault}{m}{sl}
\SetMathAlphabet{\mathsfit}{bold}{\encodingdefault}{\sfdefault}{bx}{n}

